\documentclass[final]{siamltex}
\usepackage{fullpage}
\usepackage{amsmath,amssymb}
\usepackage{xspace}
\usepackage{color}

\usepackage{url}

\newcommand\ignore[1]{}

\newcommand{\veps}{\varepsilon}

\newcommand{\ceil}[1]{\lceil{#1}\rceil}

\def\eqdef{\stackrel{{\rm def}}{=}}
\def\ala{\`a la\xspace}
\newcommand{\aset}[1]{\{#1\}}
\def\whp{with high probability\xspace}

\DeclareMathOperator{\diam}{diam}

\DeclareMathOperator{\ddim}{ddim}
\DeclareMathOperator{\MST}{MST}
\DeclareMathOperator{\OPT}{OPT}
\DeclareMathOperator{\poly}{poly}
\DeclareMathOperator{\elong}{-long}
\DeclareMathOperator{\eshort}{-short}

\newcommand{\kOPT}{$k$ \OPT}
\newtheorem{claim}[theorem]{Claim}

\newtheorem{question}[theorem]{Question}
\newtheorem{conjecture}{Conjecture}

\def\compactify{\itemsep=0pt \topsep=0pt \partopsep=0pt \parsep=0pt}

\newcounter{this-list}

\newcounter{par-list}
\newlength{\parlistlength}

\begin{document}

\title{The Traveling Salesman Problem: Low-Dimensionality Implies a Polynomial Time Approximation Scheme%
\thanks{An extended abstract of this paper appeared in Proceedings of STOC 2012.}
}

\author{
Yair Bartal\footnotemark[2]
\and
Lee-Ad Gottlieb\footnotemark[3]
\and
Robert Krauthgamer\footnotemark[4]
}

\maketitle

\renewcommand{\thefootnote}{\fnsymbol{footnote}}

\footnotetext[2]{The Hebrew University. 
Work supported in part by an Israel Science Foundation grant \#1609/11.
Email: \texttt{yair@cs.huji.ac.il}
}
\footnotetext[3]{Ariel University.
Email: \texttt{leead@ariel.ac.il}
}
\footnotetext[4]{Weizmann Institute of Science.
This work was supported in part by a US-Israel BSF grant \#2010418, and by the Citi Foundation.
Email: \texttt{robert.krauthgamer@weizmann.ac.il}
}

\renewcommand{\thefootnote}{\arabic{footnote}}

\begin{abstract}
The Traveling Salesman Problem (TSP) is among the most famous NP-hard optimization problems.
We design for this problem an algorithm that for any fixed $\veps>0$,
computes in randomized polynomial-time a $(1+\veps)$-approximation 
to the optimal tour in TSP instances that form an \emph{arbitrary} metric space
with bounded intrinsic dimension.

The celebrated results of Arora \cite{A-98} and Mitchell \cite{M-99}
prove that the above result holds in the special case of
TSP in a fixed-dimensional Euclidean space.
Thus, our algorithm demonstrates that the algorithmic tractability of metric TSP
depends on the dimensionality of the space and not on its specific geometry.
This result resolves a problem that has been open
since the quasi-polynomial time algorithm of Talwar \cite{T-04}.
\end{abstract}


\section{Introduction}

The \emph{Traveling Salesman Problem (TSP)} 
is a fundamental and extensively studied
NP-complete problem.
Indeed, numerous articles and even whole books
(\cite{Reinelt:TSP,LLRS85,Gutin02,ABCC07}) are devoted to TSP,
studying various algorithms for different families of instances.
In fact, some of the most basic techniques in combinatorial optimization
were devised to tackle TSP, including for instance cutting planes \cite{DFJ-54}.
The input for (the optimization version of) TSP
is a complete graph, whose vertex set we denote by $S=[n]$,
together with edge-weights $w(\cdot,\cdot)$ that are nonnegative and symmetric,%
\footnote{Formally, $w(x,y)=w(y,x)\ge 0$ for all $x,y\in S$.}
and the goal is to find a closed tour of $S$ of minimum (total) weight,
where a tour is simply a permutation of $S$,
i.e. it visits every vertex exactly once.

A prominent special case of TSP, called \emph{metric TSP},
is where the edge-weights satisfy the triangle inequality,%
\footnote{The triangle inequality says that
$w(x,y) \leq w(x,z)+w(z,y)$ for all $x,y,z\in S$.
}
and hence the input is simply a (finite) metric space on the point set $S=[n]$.
The importance of this variant lies in the fact that edge-weights arising in many of the typical
applications naturally represent lengths and distances.
Metric TSP offers some basic structure that may be leveraged by algorithms.
In particular, Christofides \cite{C-76} designed a $1.5$--approximation,
meaning a polynomial-time algorithm that computes a tour whose weight
exceeds the optimum by a factor of at most $1.5$.
It is a long-standing open problem to improve this approximation for metric TSP,
but it is known that there exists a constant $c>1$,
for which $c$--approximation is NP-hard \cite{PY-93,PV-06,Lampis12}.

Celebrated results of Arora \cite{A-98} and Mitchell \cite{M-99}
prove that the important special case of metric TSP
where the input metric forms a Euclidean metric,
admits a PTAS.%
\footnote{PTAS, which stands for a Polynomial-Time Approximation Scheme,
means that for every \emph{fixed} $\veps>0$ there is a $(1+\veps)$--approximation.
Note that for every constant $\veps>0$, the runtime is polynomial in $n$.
}
To be more precise, these PTAS results apply to input metrics
that are finite subsets of a \emph{fixed-dimensional} Euclidean metric
(in the case of \cite{M-99}, the Euclidean plane).
Observing that these PTAS results require two separate conditions -- Euclidean
space and fixed dimensionality --
it is only natural to ask:
\begin{question}
Do TSP instances that satisfy only one of the two properties,
\emph{bounded dimension} and \emph{Euclidean metric},
admit a PTAS?
\end{question}

The bounded-dimensionality requirement turns out to be necessary,
as Trevisan \cite{Trevisan00} shows that TSP in Euclidean metrics
(of dimension $\log n$) is NP-hard to approximate within some constant $c>1$.
It is therefore not surprising that the running time of the
aforementioned PTAS is doubly-exponential in the dimension.

Eliminating the Euclidean requirement was first addressed by Talwar \cite{T-04}.
Observe that a basic premise of this question is that
the notion of dimension applies to an arbitrary (non-Euclidean) metric space.
This is indeed possible, and Talwar relied on a definition put forth
by Gupta, Krauthgamer and Lee \cite{GKL-03}
(following \cite{Assouad83,Clarkson99}):
The \emph{doubling dimension} of a (finite) metric space $S$,
denoted $\ddim(S)$, is the smallest $k>0$ such that every ball in the metric
can be covered by $2^k$ balls of half the radius.
This definition is essentially based on volume growth,
and indeed simple volume estimates imply that a $k$-dimensional Euclidean metric
has doubling dimension $\Theta(k)$.
The opposite direction, however, is not true
and in fact the family of metrics with bounded doubling dimension
is significantly larger than that of bounded-dimensional Euclidean metrics
(see \cite{Laakso00,LP01,Laakso02,GKL-03} for details).
Talwar \cite{T-04} generalized much of Arora's machinery \cite{A-98}
and showed that TSP in metrics with fixed doubling dimension admits a QPTAS.%
\footnote{QPTAS, which stands for a Quasi-Polynomial Time Approximation Scheme,
means that for every \emph{fixed} $\veps>0$ there is a $(1+\veps)$--approximation
running in quasi-polynomial time $2^{\mathrm{polylog}(n)}$.
}
But despite repeated attempts, the original goal remained open:
\begin{question} \label{q:doubling}
Does TSP in metrics of bounded doubling dimension admit a PTAS?
\end{question}

This question has fascinated researchers, see e.g.~\cite{Lee10blog},
for several reasons.
First, the existence of a QPTAS may be interpreted as evidence that a PTAS is possible.
Second, the above question accords well with a research program
that was initiated in \cite{GKL-03,KL-04,T-04},
and studies the analogy between Euclidean metrics of fixed dimension
and general metrics of fixed doubling dimension,
from the perspective of algorithmic tractability.
It has been observed that many algorithms dealing with
the former family of metrics can be adapted to deal with the latter,
see e.g.~\cite{CG-06,GR-08,ACGP10,GKK10} and references therein for recent instantiations.
Likewise, the doubling dimension has been established as a good measure of intrinsic
dimension in the theory of metric embeddings \cite{CGT08,ABN-11,GK11,BRS11}.

A natural approach to resolving Question \ref{q:doubling}
in the positive would be to embed
the original metric space in a bounded-dimensional Euclidean space
(such embeddings were studied in \cite{ABN-08,ABN-11}), and then apply the
PTAS of Arora \cite{A-98}.
While this general approach has been quite successful in resolving many other algorithmic
problems (see for example \cite{B-96}), it fails here since any such embedding must have
non-constant distortion \cite{Laakso00,LP01,Laakso02}, 
in fact $\Omega(\sqrt{\log n})$ \cite{GKL-03}. 
It appears that achieving a PTAS for arbitrary bounded-dimensional metric
spaces requires a new approach to bypass the limitations of the embedding.

\subsection{Results}

Our central contribution is a PTAS for TSP in metrics
of fixed doubling dimension.

\begin{theorem}\label{thm:main}
A $(1+\veps)$-approximation to the optimal tour of a metric TSP instance
$S$ on $n=|S|$ points can be computed by a randomized algorithm in time \mbox{}
$n^{2^{O(\ddim(S))}} \cdot 2^{ (2^{\ddim(S)}/\veps)^{O(\ddim(S))} \sqrt{\log n} }$.
\end{theorem}

The previously known running time is is quasipolynomial in $n$,
namely $2^{(\ddim(S)\cdot\veps^{-1}\log n)^{O(\ddim(S))}}$,
due to Talwar \cite[Theorem 8]{T-04}.

\subsection{Techniques}

We build upon the framework of \cite{A-98,T-04},
and introduce two main new ideas (and several more minor ones).
Our baseline is a carefully-chosen variant of Talwar's algorithm,
and as described in Section \ref{sec:hier}, it includes:
(1) a randomized hierarchical clustering (partitioning) of $S$;
(2) the introduction of portals around every cluster;
(3) slightly modifying the optimal tour (for sake of analysis only)
so that the tour is portal-respecting
(crosses every cluster only at its portals)
and has few crossings into the cluster;
(4) a dynamic program that computes a tour for each cluster
based on the tours already computed for its subclusters.

Our first new idea (in Section \ref{sec:localMST})
is to estimate the cost incurred by an optimal tour inside a ball.
Intuitively, the estimate is merely an instantiation of the well-known
$2$--approximation of TSP using the minimum spanning tree (stated as Lemma \ref{lem:opt-mst}).
But in reality, edges entering and exiting the ball interfere with this calculation,
and thus the estimate includes both multiplicative and additive error terms.

Our second new idea is to treat separately dense regions in the metric space, meaning
balls in which an optimal tour incurs a relatively large cost. 
If all regions are sparse (not dense), then we
are almost done -- in this case we use limited randomization and enumeration,
to determine the hierarchical clustering.
Specifically, we draw at random $O(\log n)$ radii-values for every cluster center, and argue that
\whp at least one of them is useful for the construction of a good partition. We then augment the
aforementioned dynamic program to search also for the correct radii-values for 
the different cluster centers. (This appears in Section \ref{sec:sparse}.)
If there is a dense region, then we can use the first idea above to find the (nearly) smallest
one. We then ``split'' the TSP instance into two portions, roughly the inside and the outside of that
dense region. The outside is solved recursively. The inside portion is nearly sparse because it
can be covered by a limited number of smaller (and thus sparse) balls, and so it can be solved
directly by our algorithm for sparse regions. Stitching the solutions for the two portions may be costly, but
since the region is dense, we can effectively charge our algorithm's cost to that of the optimum.
(This appears in Section \ref{sec:dense}.)

\subsection{Related work}

A few hardness of approximation results are known.
That general (not necessarily metric) TSP is NP-hard
follows immediately from Karp's original NP-hardness
proof for Hamiltonian cycle \cite{K-72}.
Moreover, this proof shows that TSP does not admit
any finite factor approximation in polynomial time, unless P=NP.
Papadimitriou and Yannakakis \cite{PY-93} showed that metric TSP is
hard to approximate within some constant factor $c>1$,
even if all the metric distances are either 1 or 2.
Papadimitriou and Vempala \cite{PV-06} proved that
approximating metric TSP within factor 220/219 is NP-hard,
and Lampis \cite{Lampis12} recently improved this bound to 185/184.
Papadimitriou \cite{P-77} showed that two-dimensional Euclidean TSP is NP-hard.

The runtime of Arora's algorithm \cite{A-98} was later improved in \cite{RS98},
and his geometric approach was subsequently employed
for other Euclidean problems in \cite{CL98, ARR99, CLZ02, KR07}.
Further extension of the algorithms of \cite{A-98,T-04} to
the problem of TSP with neighborhoods (under mild conditions)
include \cite{Mitchell07} and \cite{CE11}.
Chan and Gupta \cite{CG-08} gave an algorithm for TSP
that runs in sub-exponential time in a larger family of instances,
in which an alternative notion of dimension is assumed to be bounded.

\subsection{Preliminaries}\label{sec:prelim}

Recall our notation for the metric TSP instance:
$S$ denotes the set of points, $d(\cdot,\cdot)$ their pairwise distances,
$\ddim(S)$ its doubling dimension, and $n=|S|$ its size.
We fix $0<\veps<1/20$,
which determines the approximation we eventually achieve to be $1+O(\veps)$.
We may assume that $\veps>1/n$, as otherwise all our results hold trivially
--- TSP can be solved exactly in time $O(n!)$ by straightforward enumeration,
providing better approximation and faster runtime than our claimed runtime
(which is exponential in $\poly(1/\veps)$).
By arguments found in \cite[Section 2.1.1]{A-98}, 
namely a suitable scaling and moving points at most distance $\veps n$, 
we may assume that
the minimum interpoint distance in $S$ is $1$ and the diameter is $O(n/\veps) = O(n^2)$.

As usual, the metric may be viewed as the complete graph on $S$,
with edge weights corresponding to pairwise distances,
denoted $w(x,y)\eqdef d(x,y)$.
A subset of points $S'\subseteq S$ is sometimes called a \emph{cluster}.
We let $\MST(S')$ denote a minimum-weight spanning tree (breaking ties arbitrarily)
of the complete graph on $S'$.
The \emph{ball} centered at $x\in S$ with radius $R>0$ is defined as
$B(x,R) \eqdef \aset{ y\in S:\ d(x,y) \leq R }$.
We define $B^*(x,R)$ to be the edges of the complete graph on $B(x,R)$.

\paragraph{Tours}

Throughout, a \emph{tour} $T$ is a finite sequence of points;
by convention, it is undirected, and may visit a point more than once.
A \emph{transition} in $T$ is a pair of successive points in the sequence,
which may be viewed as an edge in the complete graph on $S$
(or a self-loop of zero weight, which can be eliminated if needed).
A \emph{closed} tour is defined in the natural way by adding a transition
between the last and first points in the sequence.

The \emph{weight} (or \emph{length}) of a multiset $M$ of transitions
is defined as $w(M)\eqdef \sum_{(x,y)\in M} w(x,y)$.
This notation naturally extends to a tour $T$,
by viewing $T$ as sequence of transitions,
hence $w(T)$ represents the total length of the tour $T$.

Let $\OPT(S')$ denote a minimum-length closed tour that visits all points
of $S'\subseteq S$.


\begin{lemma}\label{lem:degree}
Let $T$ be a tour that traverses some edge $e$ more than once in the same direction.
Then there exists a lighter (smaller weight) tour $T'$,
that visits all the points visited by $T$ 
and begins and ends at the same points as $T$.
Moreover, the edges of $T'$ are a subset of the edges of $T$
(though $T'$ does not necessarily traverse them in the same direction as $T$).
\end{lemma}

\begin{proof}
We will prove the case where $T$ traverses some edge $e$ exactly twice 
in a single direction; a similar proof holds for additional traversals. 
An ordering of the edges of $T$ must take the form $E_1 e E_2 e E_3$, where 
each $E_i$ is a (possibly empty) sequence of edges, and $e=(u,v)$ is traversed twice in
the same direction, say from $u$ to $v$. 
Let $\bar{E}_2$
be a backwards ordering of $E_2$, which begins at $u$ and ends at $v$. Then
$E_1 \bar{E}_2 E_3$ visits all points visited by $T$
and has the same endpoints as $T$,
but it skips two traversals of $e$, and is thus lighter than $T$.
\end{proof}

\paragraph{Doubling dimension}
Let $\lambda_S>0$ be the {\em doubling constant} of the point set $S$, the smallest value such that every
ball in $S$ can be covered by $\lambda_S$ balls of half the radius.
Recall that the {\em doubling dimension} of $S$ is
$\ddim(S)\eqdef\log_2\lambda_S \ge 1$ (assuming $|S|\ge 2$).
%
The following packing property can be demonstrated via repeated
applications of the doubling property (see e.g. \cite{GKL-03}).
\begin{lemma} (Packing Property)
\label{lem:doublpack}
Let $S'\subseteq S$ have minimum interpoint distance $\alpha>0$. Then
$$ |S'| \leq \left( \tfrac{2\diam(S')}{\alpha} \right) ^{\ddim(S)},$$
and whenever $\tfrac{\diam(S')}{\alpha} \ge 2$,
we can further bound $|S'| \leq \left(\frac{\diam(S')}{\alpha}\right)^{2\ddim(S)}$.
\end{lemma}

\paragraph{Nets}
Similar to what was described in \cite{GGN-06,KL-04}, a subset
$S' \subseteq S$ is called a \emph{$b$-net} of $S$
if it satisfies the following two properties.
\begin{enumerate} \compactify
\renewcommand{\theenumi}{(\roman{enumi})}
\item Packing: For every $u,v \in S'$ we have $d(u,v) > b$.
\item Covering: Every $v \in S$ is within distance $b$ of some point
$u \in S'$, i.e. $S\subseteq \cup_{u\in S'} B(u,b)$.
\end{enumerate}
We say that $u\in S'$ {\em covers} $v\in S$ if $d(u,v) \le b$. The
two conditions above require that the points of $S'$ be spaced out, yet
cover all points of $S$.

\paragraph{Hierarchy of nets (or point  hierarchies)}

Recall that $\diam(S)\le O(n^2)$ 
and set $L\eqdef \ceil{\log_s \diam(S)} = O(\log_s n)$
for a parameter $s\ge 4$. (Section~\ref{sec:new} will require that $s$ is roughly
$(\log n)^{1/\ddim(S)}$.)
For each $i=0,\ldots,L$,
fix $H_i\subseteq S$ to be an $s^i$-net of $S$,
called the net of \emph{level} $i$, or of \emph{scale} $s^i$.
We may assume that the nets are nested, i.e.\ $H_i\subseteq H_{i-1}$:
Having constructed $H_i$, we may initialize $H_{i-1} = H_i$,
and then greedily add to $H_{i-1}$ uncovered points of $S$ as needed \cite{KL-04}.
Notice that the bottom level $i=0$ contains all points,
and the top level $i=L$ contains only a single point.
%

\paragraph{Net-Respecting Tours}

A tour $T$ is said to be \emph{net respecting (NR)} relative to a 
given hierarchy $\{H_i\}_{i=0}^L$ and value $\veps>0$,
if for every transition in $T$, say of length $\ell$,
both of its endpoints belong to $H_i$ for $i$ such that $s^i\leq\veps\ell < s^{i+1}$.
When the hierarchy is nested, this implies that both points
belong to {\em every} net $H_j$ with $j \le i$, although we will find
it convenient to view the edge as connecting the {\em occurrences} of the endpoints
in the single level $H_i$. 
(When $\ell < \frac{1}{\veps}$, it suffices to connect $H_0$ level points; in this case
the hierarchy implicitly contains levels $H_i$ for all $i<0$, and like $H_0$ 
these nets contain all points of $S$.)
We denote by $\OPT^{NR}(S')$ an optimal (minimum length) net-respecting tour
that visits all points of $S'\subseteq S$.

\begin{lemma}\label{lem:respect}
Let $0<\veps \leq \frac{1}{8}$.
Then every tour $T$ can be converted to a net-respecting tour $T'$
with the same endpoints which visits all points visited by $T$, such that 
$$w(T')\leq (1+16\veps) w(T).$$
\end{lemma}

\begin{proof}
For every transition $(x,y)$ in $T$ do the following.
Let $x',y'$ be the $i$-level net points covering $x,y$ respectively,
where $i$ is the highest level such that $s^i\leq 2\veps d(x,y)$.
Replace transition $(x,y)$ with $(x',y')$, and also add transitions
$(x,x')$ and $(y,y')$. The total cost of the new path is 
\[
d(x,x') + d(x',y') + d(y',y)
\le [d(x,y) + 2 \cdot 2\veps d(x,y)] + 2 \cdot 2\veps d(x,y)
= (1+8\veps) d(x,y).
\]
While transition $(x',y')$ is net-respecting, transitions
$(x,x')$ and $(y,y')$ may not be. These transitions are themselves
replaced by the procedure above. This leads to a series of transition
replacements. As the replacement procedure is activated on $2^j$ 
transitions of length at most $(2 \veps)^j d(x,y)$, and recalling
that $\veps \le \frac{1}{8}$, the total additive cost is bounded by
\[
\sum_{j=0}^\infty 8\veps \cdot 2^j \cdot (2 \veps)^j d(x,y)
\le \sum_{j=0}^\infty 8\veps 2^{-j} d(x,y)
= 16 \veps d(x,y).
\]
\end{proof}

\paragraph{Spanning trees, tours, and patching}
It is well-known that the optimal tour on a set $S'$
is approximated within factor $2$ by the minimum spanning tree on $S'$.

\begin{lemma}\label{lem:opt-mst}
Let $S'\subseteq S$.
Then $w(\MST(S')) \le w(\OPT(S')) \le 2 w(\MST(S'))$.
\end{lemma}

The following lemma, due to Talwar \cite{T-04} (see also \cite{S-10}),
uses the doubling dimension to bound $w(\MST(S'))$.

\begin{lemma}\label{lem:mst}
Let $S'\subseteq S$.
Then $w(\MST(S')) \le 4 |S'|^{1-{1}/{\ddim(S)}} \cdot\diam(S')$.
\end{lemma}

The next lemma, due to \cite{A-98,T-04},
is known as the Patching Lemma for doubling spaces.
We say that a transition $(x,y)$ in a tour $T$ \emph{crosses}
a cluster $C\subseteq S$ if exactly one of $x,y$ belongs to $C$.
The point (among $x,y$) that belongs to $C$ is called a cross-point.
A tour $T$ may cross $C$ multiple times at multiple cross-points.

\begin{lemma}[Patching Lemma] \label{lem:patching}
Let $T$ be a tour that crosses a cluster $C$ at most $r$ times,
at cross-points $\hat{C} \subset C$.
Then there is a tour 
$T'$ with the same endpoints as $T$ which visits all points visited
by $T$, crosses $C$ at most twice, and 
$$ w(T') 
  \leq w(T) + 4w(\MST(\hat C))
  \leq w(T) + 16 r^{1-\frac{1}{\ddim(S)}} \diam(\hat C).
$$
\end{lemma}

\noindent Remarks:
The last inequality is due to Lemma \ref{lem:mst}.
We sketch below the proof of this lemma for completeness,
as it is omitted from \cite{T-04}.
We also note for later reference that the bound $w(T') \leq w(T) + 4w(\MST(C))$
follows by the same proof,
except for replacing the minimum spanning tree for $\hat C$ with one for $C$.

\begin{proof}[Sketch]
For simplicity, we shall consider only a closed tour $T$,
and omit the adaptations needed for an open tour.
Break the tour $T$ at each crossing of $C$,
and fix arbitrarily two crossings to keep (the two is because $T$ is closed).

Consider first the portions of the tour that are inside $C$.
As these portions are ``disconnected'' only at points of $\hat C$
(in fact, excluding at most two of the points),
adding a minimum spanning tree on $\hat C$
of total weight $w(\MST(\hat C))$,
results in a set of edges that is connected.
Let $C_{\mathrm odd}$ be the points in $\hat C$
that have an odd degree under the current set of edges,
and add a minimum-weight matching on $C_{\mathrm odd}$.
We claim (and will prove shortly)
that the matching's total weight is at most $w(\MST(\hat C))$.
Using the claim, the current set of edges (consisting of portions of the tour,
a spanning tree, and a matching) is both connected and has even degree at
all but two vertices (the two crossings we keep).
By Euler's theorem, these edges can be arranged as
an open tour connecting the two retained crossings.
These manipulations increase the tour length by at most $2w(\MST(\hat C))$.

The same arguments apply separately to the portions outside $C$,
Patching them as before into an open tour between the two retained crossings
increases the tour length again by at most $2w(\MST(\hat C))$.
However, the patching described above introduces edges \emph{inside} $C$,
and so we add a final step to ``shortcut'' around these edges.
This shortcut maintains all visits to vertices outside $C$ (and their order),
and does not increase the tour length (by the triangle inequality).
The lemma follows by combining the two open tours.

To prove the claim concerning the minimum-weight matching on $C_{\mathrm odd}$,
define new weights $w'$ between points in $\hat C$ as follows.
Let $w'(x,y)$ be the shortest-path distance on the tree $\MST(\hat C)$,
and observe that $w'(x,y) \ge w(x,y)$.
Thus, it suffices to upper bound a minimum-weight matching (on $C_{\mathrm odd}$) 
under the tree weights $w'$. Consider such a matching,
and view every edge in the matching as a path in the tree $\MST(\hat C)$.
These tree-paths must be disjoint,
because if two tree-paths were to use the same tree-edge,
then a simple swap would decrease the weight of the matching.
Thus, this matching's weight (under $w'$) is at most
the total weight of the tree, that is at most $w(\MST(\hat C)$.
\end{proof}

We prove here another version of the Patching Lemma,
tailored to our specific needs.

\begin{lemma} \label{lem:patching2}
Let $T$ be a tour that visits all of $S$, and suppose 
it crosses a cluster $C$ at most $r$ times, at cross-points $\hat{C} \subset C$.
Let $\aset{T_i}_{i=1}^k$ be the maximal subtours of $T$ that are entirely
inside $C$.
Then there exists a closed tour $T'$ that visits all points of $C$,
contains only edges in $\cup_i T_i$ and in $\MST(\hat{C})$, and
$$ w(T')
\leq \sum_{i=1}^k w(T_i) + 2w(\MST(\hat C)) 
\leq \sum_{i=1}^k w(T_i) + 8 r^{1-\frac{1}{\ddim(S)}} \diam(C).$$
\end{lemma}
\noindent Remark: 
The last inequality is due to Lemma \ref{lem:mst} and
$\diam(\hat C)\leq \diam(C)$.

\begin{proof}
The idea is to ``stitch'' together the different subtours $T_i$,
by constructing a suitable multigraph on the vertex-set $C$.
We initialize the multigraph to be the tours $T_i$.
The endpoints of these tours are all included in the cross-points $\hat C$, 
so by adding to our multigraph the edges of $\MST(\hat C)$,
we make sure the multigraph is connected.
Let $C_{\mathrm{odd}}\subseteq C$ be the vertices of odd degree in our multigraph,
and observe that $C_{\mathrm{odd}}\subseteq \hat C$ 
because all vertices in $C\setminus \hat C$ have an even degree in $\cup_i T_i$
and degree zero in $\MST(\hat C)$.
Now add to our multigraph a minimum-weight perfect matching 
$M_{\mathrm{odd}}$ on $C_{\mathrm{odd}}$.
By well-known arguments which date back to Christofides \cite{C-76},
\[
  w(M_{\mathrm{odd}}) 
  \leq \tfrac12 w(\OPT(C_{\mathrm{odd}})) 
  \leq \tfrac12 w(\OPT(\hat C)) 
  \leq w(\MST(\hat C)),
\]
and thus the total edge-weight in our multigraph is at most
$\sum_{i=1}^k w(T_i) + 2w(\MST(\hat C))$.
Furthermore, the multigraph is Eulerian -- connected with even degrees --
and thus admits a closed tour $T'$
that visits all points visited by $\cup_i T_i$, and hence all points of $C$,
and whose weight $w(T')$ is bounded as desired.
\end{proof}

Note that the previous lemma does not address connecting the tour 
segments outside the cluster. This can be done via the minimum spanning
tree, which adds an additional weight of 
$2w(\MST(\hat C)) \leq 8 r^{1-\frac{1}{\ddim(S)}} \diam(C)$
to the final tour.





\paragraph{Exponential distribution}
In the construction of our hierarchy (Sections \ref{sec:hier} and \ref{sec:new}), 
we will need to create a ball centered at a point $u$, 
with a random radius chosen according to the exponential distribution.
Having fixed some value $a$, the density function of this distribution can take the form: 
$$f(r) = \frac{2^{8\ddim(S)}}{1-2^{-8\ddim(S)}} 
\cdot \frac{8 \ddim(S) \ln 2}{a} \cdot 2^{-\frac{8\ddim(S)}{a} r}$$
for $r \in [a,2a]$, and 0 for all other values of $r$, see \cite{ABN-11}.


\subsection{Local behavior of optimal tour}\label{sec:localMST}

We next show that the weight of the optimal net-respecting tour inside some
neighborhood can be approximated using a minimum spanning tree of points in that neighborhood.


\begin{lemma}\label{lem:mst-opt}
Let $\OPT^{NR}(S)$ be an optimal net-respecting tour visiting all points in $S$ 
(for $0 < \veps \le \frac{1}{6}$ and $s \ge 6$).
Then for all $u \in S$ and any radius $R>0$,
\begin{enumerate} \compactify
\renewcommand{\theenumi}{(\roman{enumi})}
\item \label{it:mst-opt1}
  $w(\OPT^{NR}(S) \cap {B^*(u,R)})
  \leq 6(1+16\veps)\cdot w(\MST(B(u,R)))$.
\item \label{it:mst-opt2}
  $w(\OPT^{NR}(S) \cap {B^*(u,4R)})
  \geq w(\MST(B(u,R)))
  - (s/\veps)^{2\ddim(S)}R$.
\end{enumerate}
\end{lemma}

\begin{proof}
We show that if \ref{it:mst-opt1} does not hold, we can modify the tour to
reduce its weight, which then contradicts the assumption that the tour is optimal.
Applying the Patching Lemma (Lemma \ref{lem:patching}, with the subsequent remark)
to the tour $\OPT^{NR}(S)$ with respect to the cluster $B(u,R)$,
we get a modified tour which visits all of $S$
and crosses that cluster at most twice,
while increasing the tour's length by at most $4w(\MST(B(u,R)))$.
Now replace the portion of this tour inside the cluster
with a tour that is derived from an MST of $B(u,R)$,
and thus adds total length of at most $2w(\MST(B(u,R)))$ (Lemma \ref{lem:mst}).
Finally, convert the newly added edges to be net-respecting (Lemma \ref{lem:respect});
This entire process first removes from the tour a total length
of $w(\OPT^{NR}(S)\cap {B^*(u,R)})$,
while adding a total length of at most $6(1+16\veps)\cdot w(\MST(B(u,R)))$.
Part \ref{it:mst-opt1} follows from the optimality of $\OPT^{NR}(S)$.

To prove part \ref{it:mst-opt2}, consider a tour $\OPT^{NR}(S)$.
Ball $B(u,R)$ partitions the tour into subtours $T_1,T_2,T_3,\ldots$,
where $T_k$ for odd $k$ contains only edges in $B^*(u,R)$ and 
$T_{k+1}$ contains only edges not in $B^*(u,R)$. Note that the first and
last points in $T_{k+1}$ must be in $B(u,R)$.
By definition, $w(\OPT^{NR}(S)) = \sum_k w(T_k)$.

We will now construct a connected graph $G$ whose edges are all in 
$B^*(u,4R)$ and which spans all points of $B(u,R)$, and use
a charging argument to bound its weight against 
$w(\OPT^{NR}(S) \cap {B^*(u,4R)})$. 
First, add to $G$ all subtours
inside $B^*(u,R)$ -- that is, $T_k$ for odd $k$. The cost of these edges
of $G$ are charged to the contribution of $T_k$ to $\OPT^{NR}(S) \cap {B^*(u,4R)}$.
Now consider subtours $T_{k+1}$. If $T_{k+1}$ visits only points inside $B(u,4R)$, 
then add $T_{k+1}$ to $G$, and the cost of these edges
of $G$ are charged to the contribution of $T_{k+1}$ to $\OPT^{NR}(S) \cap {B^*(u,4R)}$.

If $T_{k+1}$ exits $B(u,4R)$, then consider two more cases:
(i) If $T_{k+1}$ touches a point of the annulus $v \in B(u,4R)\setminus B(u,3R)$ 
before its initial exit from $B(u,4R)$ or after its final entrance into $B(u,4R)$,
then add to $G$ an edge connecting the first and last points of $T_{k+1}$.
Since the endpoints of $T_{k+1}$ are in $B(u,R)$, the added edge has weight
at most $2R$. Since $T_{k+1}$ connects one of its endpoints to a point in
the annulus, we have that $w(T_{k+1} \cap {B^*(u,4R)}) \ge 3R-R = 2R$, so
the added edge can be charged to the contribution of 
$T_{k+1}$ to $\OPT^{NR}(S) \cap {B^*(u,4R)}$.
(ii) If $T_{k+1}$'s first exit and final entrance into $B(u,4R)$ are from points
not in the annulus, then we add to $G$ an edge connecting the exit and 
entry points in $B(u,3R)$. Now, since these cross-points are in $B(u,3R)$, the edges
crossing $B(u,4R)$ have length at least $R$.
Let $i$ be the value for which $s^i \le \veps R < s^{i+1}$;
by the net-respecting property, these cross-points must belong to an $s^i$-net.
Since by Lemma \ref{lem:doublpack} the number of $s^i$-net points in $B(u,3R)$ is at most 
$(\tfrac{2 \cdot 6R}{\veps R/s})^{\ddim(S)} < \frac{1}{3} (s/\veps)^{2\ddim(S)}$,
the cost of adding edges connecting {\em all} $s^i$-net points in $B(u,3R)$ is at most
$3R \cdot \frac{1}{3} (s/\veps)^{2\ddim(S)} = (s/\veps)^{2\ddim(S)}R$,
from which the Lemma follows.
\end{proof}

\section{TSP via hierarchical clustering (Arora and Talwar)}\label{sec:hier}

As an exposition to our PTAS, we review a variant of
the algorithm of Talwar \cite{T-04} (and in turn Arora \cite{A-98}),
which uses hierarchical clustering
to compute a $(1+\veps)$-approximate tour in quasi-polynomial time. Recall that we may assume 
that the instance of TSP is a set $S$ with minimum interpoint distance 1 and diameter $O(n/\veps)=O(n^2)$.
The construction uses a hierarchy of nets as described above. We first introduce the single-scale partition invoked by
the algorithm -- i.e., a partition which functions separately on each hierarchical level. This partition follows the same 
framework used in \cite{B-96,B-98,FRT-03, GKL-03,ABN-11}, and is slightly
different from the one that appeared in \cite{T-04} in that it uses the exponential distribution.

\paragraph{Single-scale probabilistic partition}
Fix a set $S'\subseteq S$ to be partitioned.
Fix a level $i$, and impose
an arbitrary ordering $\pi$ on the points of the $s^i$-net $H_i \subseteq S$.
The clusters are formed one by one following the ordering $\pi$.
Each point of $H_i$ constitutes a cluster
center. With each net-point $u \in H_i$ we associate a random radius $h_u \in [s^i,2s^i]$ from an
exponential distribution.
The ball $B(u,h_u)$ constitutes a new fixed cluster of $S'$, and
then the process continues to form the rest of the clusters.
The boundary of $u$'s cluster
is determined only by the ordering imposed by $\pi$, and by the balls associated with cluster centers
at distance at most $4s^i$ from $u$.
By the packing property (Lemma \ref{lem:doublpack}), there are at most $2^{3\ddim(S)}$ such cluster centers.

The next claim follows from \cite{ABN-11}.

\begin{claim}\label{clm:prob}
For every $u,v \in S' \subseteq S$, the probability that
the single-scale probabilistic partition assigns $u$ and $v$ to
different clusters (they are {\em cut}) is at most
$\frac{c' \ddim(S) d(u,v)}{s^i}$ for some absolute constant
$c'>0$.
\end{claim}

\paragraph{Hierarchical clustering}
To create the hierarchical clustering, we first choose a single-scale partition for the top level $L$.
As described above, each net point chooses a radius in the range $[s^L,2s^L]$,
and then every point in $S$ is assigned to the first ball in $\pi$ that covers it. 
For the next hierarchical level $L-1$, we take 
each $L$-level cluster separately,
and build for its points a new partition with random radius in the range $[s^{L-1},2s^{L-1}]$. 
The construction continues recursively until level 0, the
bottom level. Note that 
each cluster has $s^{O(\ddim(S))}$ child clusters.

It follows that an $i$-level cluster is ultimately formed by independent single-scale 
partitions acting on all levels $i$ and higher. It further follows that for any point pair
$u$ and $v$, the probability that the pair are {\em cut} at level $i$ (found in different 
$i$-level hierarchical clusters) is bounded
by the sum of the probabilities that they are cut by each single-scale partition acting
on a level $i$ or higher, that is
$\sum_{j=i}^L \frac{c'\ddim(S) d(u,v)}{s^j} = O \left( \frac{\ddim(S) d(u,v)}{s^i}
\right)$.

\paragraph{TSP algorithm and analysis}
The dynamic programming TSP algorithm functions on the hierarchical clustering above.
A tour is $(m,r)$-light with respect to a fixed hierarchical partition if it crosses each
$i$-level cluster at most $r$ times, and only at a set of $m$ predetermined points, called
the {\em portals}. Following \cite{T-04}, we define the $m$ portals to be the $\frac{s^i}{M}$-net 
points in the cluster, for some value $M$ to be fixed below\footnote{In the event that the 
hierarchical cluster does not include the $\frac{s^i}{M}$-net points that cover the cluster
points, we can always add to the cluster copies of the net points. In this case, the added points
function as portals for the cluster, but do not necessarily need to be covered by a tour for the
cluster.}. Recall that the diameter of an $i$-level cluster is at most $4s^i$, and so
it follows from Lemma \ref{lem:doublpack} that $m \leq (8M)^{\ddim(S)}$. 
Throughout this section, we will take $s = 6$ 
(the minimum admissible value of $s$ in Lemma \ref{lem:mst-opt}).

An optimal $(m,r)$-light tour for the hierarchical clustering can be computed by
dynamic programming as follows: Consider a cluster $C$. Any valid $(m,r)$-light tour
crosses $C$ at most $r$ times and only at portals, so the restriction of the path to $C$
consists of at most $r$ paths
starting and ending at portals. A {\em configuration} is a multiset of $r$ or fewer
portals partitioned into pairs (each representing an entry/exit pair). A single
portal may appear more than once in the configuration if the tour crosses it multiple
times, but each instance counts towards $r$. A cluster has $m$ portals, so there are
no more than $m^{r}$ possible configurations. Now, assuming inductively that optimal $(m,r)$-light tours
have already been computed for all configurations for all $s^{O(\ddim(S))}$ children of $C$
under the hierarchical clustering, the optimal $(m,r)$-light tour for each possible 
configuration of $C$ can be
computed by a brute-force algorithm: Since the $(m,r)$-light tour of each child
cluster enters and exits via a portal, we can ``stitch'' together the child tours
through the child portals. For each fixed configuration of $C$ (at most $m^r$
possible configurations), we consider all possible child configurations ($m^{s^{O(\ddim(S))} r}$). 
Having fixed a configuration for every child cluster, we have
$s^{O(\ddim(S))} r$ candidate child portals where the tour may cross. Since
each child portal may be connected to one of $s^{O(\ddim(S))} r$ other
candidate child portals, all possible graphs connecting these portals can be
enumerated in time bounded by $\left( s^{O(\ddim(S))} r \right)^{s^{O(\ddim(S))} r}$.
Below we will choose a value for $r$ satisfying
$r = s^{\omega(\ddim(S))}$, so we can bound the previous term by 
$r^{s^{O(\ddim(S))}r}$.
For each parent configuration, we choose the valid graph with the least cost 
tour. The total runtime is $(m r)^{s^{O(\ddim(S))} r} = m^{s^{O(\ddim(S))} r}$.

Crucially, it follows from \cite{T-04} that with constant probability, the
hierarchical clustering for $S$ admits an $(m,r)$-light tour with weight at most
$(1+\veps)\OPT(S)$. Define $M$ to be the smallest power of $s$ that is greater or equal
to $\frac{\ddim(S) L}{\veps}$ -- that is,
$\frac{M}{s} < \frac{\ddim(S) L}{\veps} \le M$,
and so 
$m 
\le (8M)^{\ddim(S)} 
\le (8s \ddim(S) L / \veps)^{\ddim(S)}$.
(Recall that $L = \log_sn$.) 
Set $r$ equal to $m$. 
The proof proceeds as in \cite{A-98}, by showing that an
optimal tour can be slightly modified to observe the $(m,r)$-light property. The cost of modifying
the tour is charged to the tour's edges, and the analysis shows that the cost charged
to each edge is small. Briefly, the probability that an edge $e = (u,v)$ is cut by
the $i$-level partition is bounded by $\frac{c' d(u,v) \ddim(S)}{s^i}$. If the edge is
cut, it is rerouted through $\frac{s^i}{M}$-net points, at an additive cost
(increase in tour length) of $\frac{4s^i}{M}$. Hence, the expected cost of rerouting $e$
due to a cut at level $i$ is $\frac{c' d(u,v) \ddim(S)}{s^i} \cdot \frac{4s^i}{M} = O
\left( \frac{\veps d(u,v)}{L} \right)$, and the expected cost of rerouting $e$
due to a cut in any of $O \left( L \right)$ levels is $O(\veps\cdot d(u,v))$.
The previous step ensures that all edges crossing the cluster are incident on valid
portals. 

Now, if the optimal tour crosses an $i$-level cluster some $\tilde{r}\ge r$ times, 
the number of crossing must be decreased. In this event, the tour
is patched via the minimum spanning tree on the cross points (as in Lemma
\ref{lem:patching}). The cost is charged to the edges participating in the patching,
at a per edge cost of $O \left( \frac{s^i \tilde{r}^{1-{1}/{\ddim(S)}}}{\tilde{r}} \right) = O
\left( \frac{s^i \veps}{\ddim(S) L} \right)$. But an edge participates in a
patching only if it is cut (which happens with the probability stated above), and hence the
expected charged cost to $e$ due to patchings at one level is $ O \left( \frac{d(u,v)
\ddim(S)}{s^i} \cdot \frac{s^i \veps}{\ddim(S) L} \right) = O \left(
\frac{\veps \cdot d(u,v)}{L} \right)$, and due to patchings for all $L$ levels is
$O(\veps \cdot d(u,v))$. The values for $m$ and $r$ imply that the algorithm above
runs in quasi-polynomial time 
$m^{s^{O(\ddim(S))} r}
= 2^{(\frac{L}{\veps})^{\tilde{O}(\ddim(S))}}
= 2^{(\frac{\log n}{\veps})^{\tilde{O}(\ddim(S))}}.$


\paragraph{Runtime bottleneck} 
In closing this section, we will elaborate on why the above
algorithm does not achieve a PTAS. The runtime is directly affected by the dependence of $r$ on $L$, which causes
the term $L^{O(\ddim(S))} = (\log_sn)^{O(\ddim(S))}$ to appear in the exponent. 
The dependence of $r$ on $L$ itself stems from the fact that the probability of a pair to be cut 
by each single-scale partition is calculated
separately, and then these probabilities are summed over $L$ levels, resulting in a term $L$ appearing in the summation. 
Indeed, in the hierarchical clustering employed by the algorithm,
the event that edge $e$ is cut by an $i$-level single-scale partition, and
by no other single-scale partition, is $\Theta \left( \frac{\ddim(S)}{s^i}
\right)$. Hence, the expected cost of participating in an $i$-level patching is
mostly independent of the expected cost of participating in a $j$-level patching for
all $i \ne j$, and so a term of $L$ must appear in $r$. This is precisely the
reason why the analysis presented by Talwar \cite{T-04} does not achieve a PTAS for
metric TSP.

\section{Obtaining a PTAS}\label{sec:new}

In this section, we prove Theorem~\ref{thm:main}, the central contribution of this paper.
Henceforth, we assert the conditions of Lemma \ref{lem:mst-opt}, namely
$s \ge 6$ and $0 < \veps \le \frac{1}{6}$.

Our algorithm for TSP is an adaptation of the one employed by Talwar \cite{T-04}. His
algorithm requires a hierarchical partition, yet we cannot directly employ the
partition of Section \ref{sec:hier}. As mentioned above, that 
clustering essentially decides the cluster
assignment for each level separately, and hence we cannot successfully invoke the
analysis of \cite{A-98} to bound the expected cost of patchings per level. 

Instead, we will employ a modified version of the above partition, and analyze its
performance on net-respecting tours. We will show that
if a tour obeys some edge-sparsity property, then it admits an $(m,r)$-light tour on
a hierarchy very similar to the one above. Crucially, the edge-sparsity property
allows us to achieve $r = O((\log n)^c)$ for a small constant $c<1$, which implies a
polynomial runtime. (Although we fix the value of $c$ in the analysis, it can in
fact be taken as an arbitrarily small constant.) This partition can be found by a
``brute-force'' version of the above dynamic programming algorithm. We then show that
if the tour has edge-dense areas, then we can segment $S$ into sparse pieces, and
solve TSP separately on each.

In what follows, we will mostly consider net-respecting tours. Note that for an optimal
tour which is not net-respecting, we may impose the assumption that no point of $S$ is 
visited more than once: If the optimal tour visits a point $v$ more than once,
we may shortcut around $v$ by directly connecting its antecedent and successor points 
in the tour. However, this shortcutting is not always possible for optimal net-respecting 
tours, since connecting the antecedent and successor points may violate the net-respecting
property of the tour. Indeed, an $i$-level net-point $v$ may possess links to 
$\veps^{-O(\ddim(S))}$
$j$-level netpoints at {\em each level} $j \le i$. To address this issue, we will consider each
occurence of $v$ in the hierarchy to be a separate copy of $v$: The copy of $v$
in $H_j$ is connected to at most $\veps^{-O(\ddim(S))}$ other net-points of $H_j$. We will
also connect the copy of $v$ in $H_j$ to copies of $v$ in $H_{j-1}$ and $H_{j+1}$ (if applicable),
via edges of infinitesimally small length. 
(We shall assume that these edges are never cut, and so do not figure into the cut analysis.)
By Lemma \ref{lem:degree}, an optimal net-respecting tour
traverses each of these edges at most twice. We consider an $i$ level partition to cut
only edges incident on copies in levels $j \le i$ -- the longer edges are only cut by
higher level partitions.


\subsection{An algorithm for sparse tours}\label{sec:sparse}

In this section we show the following: For any fixed hierarchy, if there exists some 
net-respecting tour $T$ whose edges obey a specific edge-sparsity condition, then there 
exists some clustering on the hierarchy which 
supports an $(m,r)$-light tour $T'$ with low weight $w(T') \le (1+\veps) w(T)$ 
(for favorable values of $m,r$, see Lemma \ref{lem:sparse-admit}). Further, we can
find this hierarchical clustering and the tour $T$ in polynomial time (Lemma \ref{lem:sparse-alg}).
Later in Section \ref{sec:dense}, we will show that $S$ can always be broken down into subsets 
which admit edge-sparse tours.

A tour $T$ is said to be \emph{$q$-sparse}
with respect to a hierarchy $H_1,\ldots,H_L$
if for all $i\in[L]$ and $u \in H_i$,
the edges of $T$ fully contained inside the ball $B(u,3 s^i)$ have weight
$w(T \cap {B^*(u,3 s^i)}) \leq qs^i$.
The ball $B^*(u,3 s^i)$ is said to be $q$-sparse with respect
to the tour.

Suppose that an oracle had informed us that $S$ admits a net-respecting tour that is a
$(1+\veps)$-approximation to $\OPT(S)$ and is $q$-sparse.
(An oracle with a similar capability is presented in Section
\ref{sec:dense}, for 
some values of $q$.)
Then we can prove the following lemma.
(Recall that $c'$ is the constant appearing in Claim \ref{clm:prob}.)

\begin{lemma}\label{lem:sparse-admit}
Suppose $S$ admits a net-respecting $q$-sparse tour $T$.
Then there exists a hierarchical clustering for $S$ which admits an
$(m,r)$-light tour $T'$ with $w(T') \le (1+\veps)w(T)$ for
$$m := \left(8 \log_s n \cdot s \ddim(S)/\veps \right)^{\ddim(S)}$$
and
$$ r=r(q) := 18q \cdot 2^{6\ddim(S)} \ddim(S) \log_s \log n
  + \left( {2c' \ddim(S)}/{\veps} \right)^{\ddim(S)}
  + (4s/\veps)^{2\ddim(S)}.$$
\end{lemma}

We remark that the tour $T'$ need not be net-respecting.

\begin{proof}
The proof proceeds in three steps. In Step 1, we show how to construct the 
hierarchical clustering. We then prove the existence of $(m,r)$-light tour
$T'$ by showing that $T$ can be modified to cross the clusters only
at portal points (Step 2), and to cross the portal points only $r$ times
(Step 3).

\paragraph{\bf Step 1}
Fix $T$.
The hierarchical clustering closely follows the description
from Section \ref{sec:hier}, with the only difference being that the
cluster radii are chosen a little more carefully.
Consider a net-point $u \in H_j$.
Let $E_{j\eshort}$ be the edges of $T$ of length at most $s^j$, and let 
$\tilde{E}_{j\eshort} \subseteq E_{j\eshort}$ include only edges of
$T$ with at least one endpoint inside $B(u,2s^j)$. 
As a consequence of the $q$-sparsity of the ball $B(u,3s^j)$, we have that
$\sum_{e \in \tilde{E}_{j\eshort}} w(e) \le qs^j$.

Recall that we wish to assign $u$ a random radius $h_u \in [s^j,2s^j]$.
Let $V$ include all values in the (continuous) range $[s^j,2s^j]$
which cut fewer than $9q 2^{3\ddim(S)}\ddim(S)$ edges of $\tilde{E}_{j\eshort}$. 
Since $B(u,2s^j)$ is $q$-sparse, 
the sum of edge lengths in $\tilde{E}_{j\eshort}$ is at most $qs^j$,
and so a simple averaging argument gives that less than a fraction
$\frac{1}{9 \cdot 2^{3\ddim(S)}\ddim(S)}$ of radii in $[s^j,2s^j]$
intersect more than $9q \cdot 2^{3\ddim(S)}\ddim(S)$ edges of $\tilde{E}_{j\eshort}$.
We choose $h_u$ randomly from an exponential distribution on $[s^j,2s^j]$
and resample until finding a $h_u \in V$. 
Recalling that the density function $f(\cdot)$ of the exponential distribution
is decreasing, and setting parameter $a=s^j$, 
the probability that a sampled radius is invalid is less than
$f(a) \cdot \frac{a}{9 \cdot 2^{3\ddim(S)}\ddim(S)} < 2^{-3\ddim(S)} \ln 2$.

The rest of the clustering is done exactly as before by iterating over centers.
Note that knowledge of $T$ was necessary to
determine which radii are valid choices for $h_u$.

\paragraph{\bf Step 2}
We now analyze the expected cost of converting the tour $T$ 
to cross every cluster only through its $m$ cluster portals. 
Consider some $j$-level cluster $C$, and recall that $C$ is formed by combining a sequence of single-scale
partitions in levels $i \ge j$.
Let $M$ be the smallest power of $s$ at least $\frac{\ddim(S) \log_sn}{\veps}$, 
and as above we define the $m$ portals to be $\frac{s^j}{M}$-net points in the cluster.
So the number of portals is at most
$(8M)^{\ddim(S)} \le \left( \frac{8s \log_sn \ddim(S)}{\veps} \right)^{\ddim(S)}$ 
as required.

Recall from above that $E_{j\eshort}$ is the set of edges of $T$ of length at most $s^j$, 
and similarly define $E_{j\elong}$ to be the edges of $T$ of length greater than $s^j$.
Since $T$ is net-respecting, the edges of $E_{j\elong}$ must all incident on $s^k$-net points
(or higher level points) for $s^k \le \veps s^j < s^{k+1}$. Since $M$ is a power of $s$ and
$\veps s^j > \frac{s^j}{M}$, we conclude that $s^k \ge \frac{s^j}{M}$.
Hence these long edges cross $C$ at a subset of the $\frac{s^j}{M}$-net points,
and no further action is required.

We turn to edges $E_{j\eshort}$. Claim \ref{clm:prob} asserted that the
probability that a given edge $e = (u,v) \in T$ is cut by 
an $i$-level single-scale partition is
bounded by $\frac{c' d(u,v) \ddim(S)}{s^i}$.
We show that the probability that $e$ is cut {\em conditioned} on the choosing
only valid radii (those belonging to $V$) is at most $\frac{16c' d(u,v) \ddim(S)}{s^i}$:
An $i$-level ball cutting $e$ is within distance $2s^i$ of an endpoint of $e$, and so
by Lemma \ref{lem:doublpack} at most
$b = 2(2 \cdot 4)^{\ddim(S)} = 2 \cdot 2^{3\ddim(S)}$ balls may cut $e$. The partition imposed an
ordering on these balls. Let $E_\ell$ be the event that the radius of the $\ell$-th
ball covers exactly one endpoint of $e$, and $F_\ell$ be the event that it covers neither.
Then the probability that $e$ is cut is exactly 
$\Pr[E_1] 
+ \Pr[E_2] \Pr[F_1] 
+ \Pr[E_3] \Pr[F_1] \Pr[F_2] 
+ \ldots 
+ \Pr[E_b] \Pi_{i=1}^{b-1} \Pr[F_i]$.
Conditioning on choosing a valid radius increases this sum by at most a factor
$(1-2^{-3\ddim(S)} \ln 2)^{-b} < e^{2b2^{-3\ddim(S)} \ln 2} = 16$, as claimed.

If $e \in E_{j\eshort}$ is cut, we reroute it through $\frac{s^i}{M}$-net points, increasing the tour by
at most $\frac{4s^i}{M}$.
So the expected cost of moving $e$ due to a cut in level $i$ is at most
$\frac{16c' d(u,v) \ddim(S)}{s^i} \cdot \frac{4s^i}{M}
= O \left( \frac{\veps d(u,v)}{\log_s n} \right)$,
and the expected cost of moving $e$ due to cuts in all $L=O \left( \log_s n
\right)$ levels is $O(\veps d(u,v))$.


\paragraph{\bf Step 3}
Finally, we turn to the analysis of reducing the number of utilized cross-points to $r$ via patching. 
First consider edges of $E_{j\elong}$. As explained above, these are incident on
$s^k$-net points (or higher level net points) for $s^k \le \veps s^j < s^{k+1}$.
By the packing property, these cross-points account for at most 
$(4s^j/s^k)^{\ddim(S)} < (4s/\veps)^{\ddim(S)}$ active portals. 
Since a cluster may have at most $(4s)^{\ddim(S)}$ sibling clusters, and a tour may
traverse each edge at most twice (Lemma \ref{lem:degree}), 
these acount for at most 
$2(4s)^{\ddim(S)} (4s/\veps)^{\ddim(S)} < (4s/\veps)^{2\ddim(S)}$ crossings.
We can afford to retain all these crossings.

We turn to the short edges of $E_{j\eshort}$. 
Consider some $j$-level cluster $C$ centered at $u$, and recall that $C$ is formed by 
choosing a valid radius $h_u \in [s^j,2s^j]$, and
further combining a sequence of single-scale partitions in levels $i \ge j$. 
Since the radius of each
$(i \ge j)$-level single-scale partition is chosen from $V$, 
it cuts at most $9 q \ddim(S)$ edges of length at most $s^j$.
Further, edges crossing $C$ could have actually been cut by any $i$-level ball whose center is within distance
$h_u + 2s^i \le 4 s^i$ from the center of $C$; there are at most $2^{3\ddim(S)}$ such balls at each level
$i$. It follows that the number of edges in $\tilde{E}_{j\eshort}$ cut
by each $i$-level single-scale partition is at most $2^{3\ddim(S)} \cdot 9q \cdot 2^{3\ddim(S)}\ddim(S)$. Set 
\[
r'=r'(q):= \max \left \{ 2^{6 \ddim(S)} \cdot 18 q \ddim(S) \log_s \log n ,
                  \left( \frac{2c' \ddim(S)}{\veps} \right)^{\ddim(S)} \right \}
\]
and consider the case where at least $r'$ edges of $\tilde{E}_{j\eshort}$ are cut.
Then at least $r'/2$ edges of $\tilde{E}_{j\eshort}$ must have been
cut by balls in levels $i \ge j+ \log_s \log n$,
and we can charge a patching for $j$-level cluster $C$ only to these short edges.

Now, if more than $r'$ short edges cross $C$, the tour is patched via the minimum
spanning tree (\ala Lemma \ref{lem:patching}), at a per edge cost of
$O \left( \frac{s^j {r'}^{1-{1}/{\ddim(S)}}}{r'/2} \right)
= O \left( \frac{s^j \veps}{\ddim(S)} \right)$.
Recall though that the edges charged for this patching are edges cut by balls at levels $j+\log_s \log n$ or
higher. It follows that the expected cost to edge $(u,v)$ due to a patching for $j$-level
cluster $C$ is
$ O \left( \frac{d(u,v) \ddim(S)}{s^{j+\log_s \log n}}
\cdot \frac{s^j \veps}{\ddim(S)} \right)
= O \left( \frac{\veps d(u,v)}{\log n} \right)$,
and due to patchings at all levels is $O(\veps d(u,v))$.
This concludes the analysis for the short edges, and together with the long edges
the total number of cross-points is at most
$r:= r' + (4s/\veps)^{\ddim(S)}$.
\end{proof}

Let $s=(\log n)^{1/(c''\ddim(S))}$ for some constant $c'' \geq 32$.
We can now provide an efficient algorithm to find a tour with the guarantees of the 
last lemma.

\begin{lemma}\label{lem:sparse-alg}
If $S$ admits a net-respecting $q$-sparse tour $T$, then there exists a randomized algorithm that, with constant probability,
finds a tour $T'$ with $w(T') \le (1+\veps)w(T)$ in time
$n^{O(2^{4\ddim(S)})} \cdot 2^{O(q (\ddim(S)/\veps)^{3\ddim(S)} \sqrt[4]{\log n})}$.
\end{lemma}

\begin{proof}
If we could compute a hierarchical clustering that realizes Lemma \ref{lem:sparse-admit}, then the standard
dynamic program from Section \ref{sec:hier} would give a tour for $S$ satisfying Lemma \ref{lem:sparse-alg}. However, we cannot
compute this hierarchical clustering, since we do not have access to $T$ and cannot know which radii are valid choices
for $h_u$. Instead, we present a dynamic program that guesses the proper value of $h_u$. 
Recall that the exponential distribution of \cite{ABN-11} implies that a random guess for the value of $h_u$ 
is a valid value with probability at least $1/2$.
Hence, $O(\log n)$ independent random choices ensure that at least one choice for $h_u$ is
valid with probability $1-\frac{1}{n^2}$, which by a union bound implies
that with constant probability,
for each net-point at least one of its $O(\log n)$ choices is valid.

We begin by fixing $O(\log n)$ random radius choices for each net-point. 
Now consider some $j$-level cluster $C$ centered at $u \in H_j$. $C$ is formed by cuts from neighboring balls
in levels $j$ and above, and we wish to enumerate all possible {\em formations} of $C$:
Recall that we make $O(\log n)$ random choices for $h_u \in [s^j,2s^j]$.
Further, since for all $i \ge j$, $u$ is within distance $h_u + 2s^i < 4s^i$ of $2^{3\ddim(S)}$ other $s^i$-net
points whose radii may cut $C$, and we guess $O(\log n)$ radii for each of these net-point, $C$ may be cut
in $(O(\log n))^{2^{3\ddim(S)}}$ different ways by the $i$-level partition. Since $C$ may be cut from above in all 
$L-j$ levels, it follows that the number of possible formations for $C$ is bounded by
$(O(\log n))^{2^{3\ddim(S)} L}$.
Recall that $s = (\log n)^{1/c''\ddim(S)}$, and it follows that the number of levels
in the hierarchy is $L= O(\log_s n) = O \left( \frac{\ddim(S) \log n}{\log \log n} \right)$. 
So the number of possible formations for $C$ is bounded by 
$(O(\log n))^{2^{3\ddim(S)} L} = n^{O(2^{4\ddim(S)})}$.

Having fixed all random radii, we compute tour $T$ via a dynamic programming 
algorithm which executes an exhaustive search.
The dynamic programming table possesses a single entry for each possible portal configuration
of each possible formation of each cluster center.
So the table possesses $m^r \cdot n^{O(2^{4\ddim(S)})}$ entries.
The algorithm must compute for each entry an optimal cluster tour for the particular cluster formation
and portal configuration.
The table is filled in a bottom-up fashion, from level $0$ to level $L$, 
and the algorithm computes the entry for a $j$-level cluster by consulting the entries of its child
clusters in level $j-1$:
An $j$-level cluster $C$ has at most $s^{2\ddim(S)}$ child clusters, 
and since each cluster has $O(\log n)$ possible radii, there are 
$(O(\log n))^{s^{2\ddim(S)}}$
possible child formations. 
For each fixed child formation, we consider each portal configuration for the set of children
($(m^r)^{s^{2\ddim(S)}}$ possibilities), and consult the appropriate table entries for the cost
of the optimal child cluster tours.
We then compute the cost of connecting the child portals to form a valid tour through the portals of $C$:
This can be done by enumerating all graphs with one edge on each vertex, in at most
$\left( rs^{2\ddim(S)} \right)^{rs^{2\ddim(S)}}$ different ways.

Note that (assuming sufficiently large $\ddim(S) = \Omega(1)$)
$$r \log m
= O(q 2^{6 \ddim(S)} \log\log n 
\cdot (s \ddim(S) /\veps)^{2\ddim(S)} 
\cdot \log\log n 
\cdot \log(1/\veps))
= O( q (\ddim(S)/\veps)^{3 \ddim(S)} s^{2 \ddim(S)} (\log \log n)^2).$$
So we can bound each of the expressions above by:
$$(m r \log n)^{O(r s^{2\ddim(S)})} 
= 2^{O(r \log m s^{2\ddim(S)})}
= 2^{O(q (\ddim(S) /\veps)^{3\ddim(S)} s^{4\ddim(S)} (\log \log n)^2)} 
= 2^{O(q (\ddim(S) /\veps)^{3\ddim(S))} \sqrt[4]{\log n})}.$$
Lemma \ref{lem:sparse-alg} follows.
\end{proof}

\ignore{
\paragraph{Comment}
For use in Section~\ref{sec:depend}, we note that the previous construction gives an alternate
(perhaps stronger) guarantee to that of Lemma \ref{lem:sparse-alg}. Let a $k$-tour of $S$ be a set of
$k$ {\em open} tours that
amongst them visit all points of $S$. The best $k$-tour is denoted as $\kOPT(S)$. The term $(m,r)$-light
with respect to $k$-tours is unchanged; each cluster may only be entered or exited via the $r$ portals, once per
portal. Similar, a $q$-sparse $k$-tour is one in which each $i$-level ball $B(u,3s^i)$ covers at most $qs^i$
edges of the $k$-tour. Then the following lemma holds.

\begin{lemma}\label{lem:sparse-kalg}
If $S$ admits a net-respecting $q$-sparse $k$-tour $T$, then there exists a randomized algorithm that
with constant probability finds a $k$-tour $T'$ with $w(T') \le (1+\veps)w(T)$ in time
$n^{O(2^{4\ddim(S)})} \cdot 2^{O(q\log q (\ddim(S)/\veps)^{4\ddim(S)} \sqrt[4]{\log n})}$.
\end{lemma}
}

\subsection{Eliminating dense areas}\label{sec:dense}

Lemmas \ref{lem:sparse-admit} and \ref{lem:sparse-alg} show that sparse tours admit efficient
hierarchical decompositions and algorithms. Here, we consider tours that have dense neighborhoods, and
show that the point set can be divided into areas with all light tours.
We then solve TSP on each subset, and join the resulting subtours into a
single tour.

Set $q:= (s/\veps)^{O(\ddim(S))} \cdot 2^{O(\ddim^2(S))}$.

\begin{lemma}\label{lem:dense}
There is a (randomized) polynomial-time algorithm that given a set $S$ (with $|S|>1$),
computes two subsets $S_1 \subset S$ and $S_2 \subsetneq S$ with $S_1 \cup S_2=S$ and $S_1 \cap S_2 \ne \emptyset$,
such that
\begin{enumerate} \compactify
\renewcommand{\theenumi}{(\alph{enumi})}
\item\label{it:s1}
$\OPT^{NR}(S_1)$ is $q'$-sparse,
for $q' = O(q \sqrt[8]{\log n})$ ; and
\item\label{it:s2}
$w(\OPT^{NR}(S_1)) + w(\OPT^{NR}(S_2)) \le w(\OPT^{NR}(S)) + \veps w(\OPT^{NR}(S_1))$.
\end{enumerate}
\end{lemma}

Before proving Lemma \ref{lem:dense}, we will demonstrate that it can be used to complete
the proof of Theorem \ref{thm:main}.

\begin{proof}[Proof of Theorem \ref{thm:main}]
Given a point set $S$, if $S$ contains a single point then we are done.
Otherwise, we use the procedure of Lemma~\ref{lem:dense} to create two instances of TSP, $S_1 \subseteq S$
and $S_2 \subset S$. $S_1$ admits a $q'$-sparse and net-respecting tour $\OPT^{NR}(S_1)$ as promised by Lemma~\ref{lem:dense}.
A tour of almost the same cost (at most $1+\veps$ factor larger) can be computed by the algorithm of Lemma \ref{lem:sparse-alg}, obtaining a tour $T_1$, where $w(T_1) \leq
(1+\veps) w(\OPT^{NR}(S_1))$. The tour for $S_2$ is solved recursively (that is $S$ is replaced by $S_2$),
obtaining a tour $T_2$. The inequality $S_1 \cap S_2 \ne \emptyset$ implies that separate tours $T_1$ and $T_2$ can be joined
together to obtain a complete tour $T$ at no additional cost.

We now prove inductively that $w(T) \leq \left(\frac{1+\veps}{1-\veps}\right) \cdot w(\OPT^{NR}(S))$.
By the induction hypothesis we have that $w(T_2) \leq \left(\frac{1+\veps}{1-\veps}\right) \cdot w(\OPT^{NR}(S_2))$.
Lemma~\ref{lem:dense} implies that
\begin{equation*}
w(T_1) \leq (1+\veps) w(\OPT^{NR}(S_1)) \leq \left(\frac{1+\veps}{1-\veps}\right) \cdot (w(\OPT^{NR}(S)) - w(\OPT^{NR}(S_2))).
\end{equation*}
Therefore
\begin{equation*}
w(T) = w(T_1) + w(T_2) \leq w(T_1) +\left(\frac{1+\veps}{1-\veps}\right) \cdot w(\OPT^{NR}(S_2)) \leq \left(\frac{1+\veps}{1-\veps}\right) \cdot w(\OPT^{NR}(S)),
\end{equation*}
proving the inductive claim.
Finally, by Lemma~\ref{lem:respect} we have that $w(T) = (1+O(\veps))\cdot w(\OPT(S))$.
The runtime follows from executing the algorithm of Lemma \ref{lem:sparse-alg} on the
$q'$-sparse sets of Lemma \ref{lem:dense}.
\end{proof}

We now return to proving Lemma \ref{lem:dense}. We require a preliminary lemma.
Define the annulus $A(v,r_1,r_2) = B(v,r_2) \setminus B(v,r_1)$, and let
$A^\ast(v,r_1,r_2)$ be the set of edges with both endpoints inside
the annulus $A(v,r_1,r_2)$.

\begin{lemma}\label{lem:predense}
For any level $i$, let $v \in S$ be a point for which
$w(\MST(B(v,s^i)))$ is maximized, and let this weight be $q^\ast s^i$.
Let $T = \OPT^{NR}(S)$. If $q^\ast \ge 6.5$ the following hold.
\begin{enumerate} \compactify
\renewcommand{\theenumi}{(\roman{enumi})}
\item
$w(\MST(B(v,13s^i))) < 2^{5\ddim(S)} \cdot q^\ast s^i$.
\item
Set $\delta \le \frac{1}{12}$. There exists a radius $h \in [12s^i, 13s^i]$ for which 
$$w(T \cap A^\ast(v,h-6\delta s^i, h+6\delta s^i)) < 144 \delta (1+16\veps) w(\MST(B(v,13 s^i))).$$
\item 
Let $h$ be as above, let $k$ satisfy $s^k \le \delta s^i < s^{k+1}$, and let $N(h)$ 
denote the set of all $k$-level points which cover points of 
$A(v,h-\delta s^i, h+\delta s^i)$. Then 
$$\sum_{u \in N(h)} w(\MST(u,s^k)) 
< 2^{5\ddim(S)} w(T \cap A^\ast(v,h-6\delta s^i, h+6\delta s^i)) 
+ (2s^2/\veps\delta)^{2\ddim(S)} s^k.$$
\end{enumerate}
\end{lemma}

\begin{proof}
To prove the first item: Lemma \ref{lem:doublpack} implies that 
$B(v,13s^i)$ can be covered by 
$(2 \cdot 13/3)^{\ddim} < 2^{4\ddim(S)}$ 
balls of radius $3s^i$ centered at points of $S$.
Note that $w(\MST(B(v,13 s^i)))$ is bounded by the cost of
constructing a minimum spanning tree inside each of these small balls
and then connecting the balls together. 
By choice of $v$, each small ball has minimum spanning tree weight at most
$q^* s^i$, so the sum of the weights of these minimum spanning trees is less than
$2^{4\ddim(S)} q^\ast s^i$.
By Lemma \ref{lem:mst}, the centers of the small balls
can be joined by an spanning tree of weight
$4(2^{4\ddim(S)})^{1-1/\ddim(S)} \cdot 26 s^i = 2^{4\ddim(S)} \cdot 6.5 s^i$.
It follows that
$w(\MST(B(v,13s^i)) < 2^{4\ddim(S)} (q^\ast + 6.5) s^i \le 2^{5\ddim(S)} \cdot q^\ast s^i$.

To prove the second item: By an averaging argument, there is a value for $h$ for which 
$T \cap A^\ast(v,h-6\delta s^i, h+6\delta s^i)$ 
contains edges of total weight at most
$12 \cdot 2\delta \cdot w(T \cap B^\ast(v, 13s^i) )$.
By Lemma~\ref{lem:mst-opt}\ref{it:mst-opt1}, 
$w(T \cap B^\ast(v, 13 s^i)) \leq 6(1+16\veps) w(\MST(B(v,13 s^i)))$.
The item follows.

To prove the third item:
By Lemma \ref{lem:mst-opt}\ref{it:mst-opt2}
for each $k$-level net point $u$ we have
$w(\MST(B(u,s^k))) \leq w(T\cap B^\ast(u,4s^k)) + (s/\veps)^{2\ddim(S)} s^{k}$.
By Lemma \ref{lem:doublpack}, the number of $k$-level net points covering the annulus is upper bounded by
$(2 \cdot 2((13+\delta)s^i + s^k )/s^k)^{\ddim(S)} 
< (2 \cdot 2(1 + 13.1s/\delta))^{\ddim(S)} 
< (2s/\delta)^{2\ddim(S)}$. Hence,
$$\sum_{u \in N(h)} w(\MST(B(u,s^k))) \leq \sum_{u \in N(h)}
\left[ w(T\cap B^\ast(u,4s^k)) \right] + (2s^2/\veps \delta)^{2\ddim(S)} s^k.$$
Now, each ball $B(u,4s^k)$ is fully contained in the larger annulus 
$A^\ast(v,h-6\delta s^i, h+6\delta s^i) )$, and intersects at most 
$(2 \cdot 16)^{\ddim(S)} = 2^{5\ddim(S)}$ other balls. So
$\sum_{u \in N(h)} w(T\cap B^\ast(u,4s^{k}) ) 
\leq 2^{5\ddim(S)} w(T\cap A^\ast(v,h-6\delta s^i, h+6\delta s^i) )$,
and the item follows.
\end{proof}

Finally, we complete the proof of Lemma \ref{lem:dense}:

\begin{proof}[Proof of Lemma \ref{lem:dense}]
Suppose first that for all level $i$ and $u \in S$, the edge-sparsity condition
$w(\MST(B(u,3s^i)) \le 2q s^i$ holds.
then we conclude by Lemma \ref{lem:mst-opt}\ref{it:mst-opt1} that $\OPT^{NR}(S)$ is 
$13q$-sparse,
and our lemma is trivial: Set $S_1=S$ and $S_2$ includes an arbitrary single point. 
Assume then that the edge-sparsity condition does not hold.
The algorithm begins by locating the lowest level $i$ for which there exists $u \in S$ such that
$w(\MST(B(u,3s^i)) > 2q s^i$, and setting $v$ to be such that $w(\MST(B(v,3s^i)))$ is maximized.
Let $q^\ast:= w(\MST(B(v,3s^i)) / s^i$, and it follows that $q^\ast > 2q$.

Fix a tour $T=\OPT^{NR}(S)$. Ideally, we would now like to choose some radius $h$,
partition $S$ into two point sets
$\tilde{S}_1 = B(v,h)$ and $\tilde{S}_2 = S\setminus\tilde{S}_1$,
and then provide tours for the two sets
whose combined weight is only slightly greater than that of $T$.
Let $\tilde{S}_i^\ast$ denote the edges of the complete graph
on the points of $\tilde{S}_i$ (for $i=1,2$); then we could bound the weight of the subtours by showing that each
$T_i = \tilde{S}_i^\ast \cap T$ (for $i=1,2$) can be made into a closed tour by
adding only a light-weight collection of edges to ``patch'' the edges of $T$ cut by the partition (as
in Lemma \ref{lem:patching2}). However, this plan may be costly
because the patchings might be expensive; for example,
a radius $h$ ball can cut many edges of $T$ which then need to be patched.
Moreover, in order to ensure that the subtours are net-respecting, we need to
augment the sets $\tilde{S}_i$ with appropriate (nearby) net points.
To solve this problem, we will show how to create sets $S_1 \supset \tilde{S}_1$ (which 
also contains some points of $\tilde{S}_2$) and $S_2 \supset \tilde{S}_2$ (which also contains some
points of $\tilde{S}_1$) for which the lemma holds.

We choose a radius $h \in [12s^i,13s^i]$ given by Lemma \ref{lem:predense}.
In what follows, we will show separately how to patch long and short edges crossing
$\tilde{S}_1$. (Similarly arguments allow for patching $\tilde{S}_2$.)

Let $E_{i\elong}$ be the set of long edges of $T$ crossing $\tilde{S}_1$, those of length more than $\delta 
s^i$ for $\delta = O(\veps/2^{10\ddim(S)})$. Since $T$ is net-respecting, these edges must cross $\tilde{S}_1$ 
at $j$-level net points, where $j$ satisfies $s^j \le \veps\delta s^i < s^{j+1}$. We will 
patch the edges of $E_{i\elong}$ crossing $\tilde{S}_1$ using the minimum spanning tree of {\em all} 
$j$-level net points covering $\tilde{S}_1$. There are $(s/\veps \delta)^{4\ddim(S)}$ such net points, and 
their net-respecting $\MST$ has weight less than $(1+16\veps)(s/\veps \delta)^{4\ddim(S)} s^i < \veps q s^i$ 
for an appropriate choice of $q$
(see Lemmas \ref{lem:respect}, \ref{lem:mst}), which bounds, up to a constant factor, the cost of patching the 
long edges crossing $\tilde{S}_1$.

We now turn to patching the shorter edges of $T$ that cross $\tilde{S}_1$. Let $E_{i\eshort}$ include edges
of length at most $\delta s^i$.
Now, these edges cross into $\tilde{S}_1$ from a set of points $V \subset \tilde{S}_2$ 
inside the annulus $A(v,h-\delta s^i,h+\delta s^i)$. Since we have shown how to patch the long edges
of $T_1$ crossing $j$-level net points,
the short edges of $T_1$ crossing $V$ will be patched by connecting
them to $k$-level net points (though not necessarily directly) where $k$ satisfies
$s^k \leq \delta s^i < s^{k+1}$.
To this end, add to $\tilde{S}_1$ copies of all $k$-level net points
which cover points in $V$ -- call this set $N(h)$ -- 
as well as all lower level points within distance $s^k$ of $N(h)$, 
and let the resulting point set be $S_1$.
We patch the short edges via the MST of the $s^k$-radius balls of points in $N(h)$, 
that is $\cup_{u \in N(h)} \MST(B(u,s^k))$. By the three items of Lemma \ref{lem:predense}, when
$q \geq \veps^{-1}(2s^2/\veps\delta)^{2\ddim(S)}$ we have that the total cost of this patching is
$$\cup_{u \in N(h)} w(\MST(B(u,s^k))) = O(\delta) \cdot 2^{10\ddim(S)} q^\ast s^i.$$
Since $\delta = O(\veps/2^{10\ddim(S)})$, this last bound is 
$O(\veps q^\ast s^i)$. We then reroute all new edges to be net-respecting 
(adding to $\tilde{S}_1$ net points of $\tilde{S}_2$ as necessary)
at a trivial cost (Lemma \ref{lem:respect}).

We conclude that the total weight of all patchings (including short and long edges) is 
$O(\veps q^\ast s^i)$.
By Lemma~\ref{lem:mst-opt}\ref{it:mst-opt2}
(and since $S_1 \supset B(v,12s^i)$) we have
$w(\OPT^{NR}(S_1)) \geq w(\MST(B(v,3 s^i))) - (s/\veps)^{2\ddim(S)} 3s^i
= q^* s^i - (s/\veps)^{2\ddim(S)} 3s^i \ge \frac{1}{2} q^\ast s^i$, 
so the patching cost is indeed $O(\veps)\cdot w(\OPT^{NR}(S_1))$.
A similar argument applies to the patchings needed for $S_2$, by adding to
$\tilde{S}_2$ all $j$-level net points of $\tilde{S}_1$, as well as all
$k$-level net points of $\tilde{S}_1$ covering points of $\tilde{S}_2$ and their balls;
the resulting set is $S_2$. This completes the proof of part \ref{it:s2} of the lemma.

We note that by construction $S_1 \cap S_2 \neq \emptyset$, and that
$S_2 \neq S$ (as implied by part \ref{it:s2} of the lemma).

We now complete the proof of part \ref{it:s1}, translating balls with light MST to light tours.
By the choice of $i$, for every $\ell<i$ and every net-point $u \in H_\ell$,
$\MST(B(u,3s^\ell)) \le 2q s^\ell$, and so by Lemma~\ref{lem:mst-opt}\ref{it:mst-opt1},
$w(\OPT^{NR}(S_1) \cap B^\ast(u,3s^\ell)) \le 6(1+16\veps) q s^\ell \le 22 q s^\ell$.
While by assumption the $\ell$-level balls have light minimum spanning trees,
this is not tree of the $i$-level ball $B(v,3s^i)$. Fix $\ell$ to be the value satisfying 
$s^\ell \le s^i < s^{\ell+1}$, and by the packing property the $i$-level ball covers 
$(4s)^{\ddim(S)}$ $\ell$-level balls. It follows that 
$w(\OPT^{NR}(S_1) \cap B^\ast(u,3s^i)) 
\le (4s)^{\ddim(S)} 22 q s^\ell 
= O(\sqrt[16]{\log n}) \cdot q s^i$, 
which completes the proof.

\end{proof}

\paragraph{Acknowledgements}
We thank Ittai Abraham, Alex Andoni, Anupam Gupta, Liam Roditty and Kunal Talwar
for helpful discussions.

{\small 
\bibliographystyle{siam}
\bibliography{bib-hier-sicomp}

\begin{thebibliography}{10}

\bibitem{ABN-08}
{\sc Ittai Abraham, Yair Bartal, and Ofer Neiman}, {\em Embedding metric spaces
  in their intrinsic dimension}, in Proceedings of the nineteenth annual
  ACM-SIAM symposium on Discrete algorithms, 2008, pp.~363--372.

\bibitem{ABN-11}
\leavevmode\vrule height 2pt depth -1.6pt width 23pt, {\em Advances in metric
  embedding theory}, Advances in Mathematics, 228 (2011), pp.~3026 -- 3126.

\bibitem{ACGP10}
{\sc Ittai Abraham, Shiri Chechik, Cyril Gavoille, and David Peleg}, {\em
  Forbidden-set distance labels for graphs of bounded doubling dimension}, in
  29th ACM SIGACT-SIGOPS symposium on Principles of distributed computing, ACM,
  2010, pp.~192--200.

\bibitem{ABCC07}
{\sc David~L. Applegate, Robert~E. Bixby, Vasek Chvatal, and William~J. Cook},
  {\em The Traveling Salesman Problem: A Computational Study (Princeton Series
  in Applied Mathematics)}, Princeton University Press, Princeton, NJ, USA,
  2007.

\bibitem{A-98}
{\sc Sanjeev Arora}, {\em Polynomial time approximation schemes for {E}uclidean
  traveling salesman and other geometric problems}, J. ACM, 45 (1998),
  pp.~753--782.

\bibitem{ARR99}
{\sc Sanjeev Arora, Prabhakar Raghavan, and Satish Rao}, {\em Approximation
  schemes for {E}uclidean k-medians and related problems}, in 30th annual ACM
  symposium on Theory of computing, ACM, 1998, pp.~106--113.

\bibitem{Assouad83}
{\sc Patrice Assouad}, {\em Plongements lipschitziens dans {${\bf R}\sp{n}$}},
  Bull. Soc. Math. France, 111 (1983), pp.~429--448.

\bibitem{B-96}
{\sc Yair Bartal}, {\em Probabilistic approximation of metric spaces and its
  algorithmic applications}, in 37th Annual Symposium on Foundations of
  Computer Science, IEEE Computer Society, 1996, pp.~184--193.

\bibitem{B-98}
\leavevmode\vrule height 2pt depth -1.6pt width 23pt, {\em On approximating
  arbitrary metrices by tree metrics}, in 30th annual ACM symposium on Theory
  of computing, ACM, 1998, pp.~161--168.

\bibitem{BRS11}
{\sc Yair Bartal, Ben Recht, and Leonard~J. Schulman}, {\em Dimensionality
  reduction: beyond the {J}ohnson-{L}indenstrauss bound}, in Proceedings of the
  Twenty-Second Annual ACM-SIAM Symposium on Discrete Algorithms, SODA '11,
  SIAM, 2011, pp.~868--887.

\bibitem{CE11}
{\sc T.-H.~Hubert Chan and Khaled~M. Elbassioni}, {\em A {QPTAS} for {TSP} with
  fat weakly disjoint neighborhoods in doubling metrics}, Discrete {\&}
  Computational Geometry, 46 (2011), pp.~704--723.

\bibitem{CG-08}
{\sc T-H.~Hubert Chan and Anupam Gupta}, {\em Approximating {TSP} on metrics
  with bounded global growth}, in 19th annual ACM-SIAM symposium on Discrete
  algorithms, SIAM, 2008, pp.~690--699.

\bibitem{CGT08}
{\sc T-H.~Hubert Chan, Anupam Gupta, and Kunal Talwar}, {\em
  Ultra-low-dimensional embeddings for doubling metrics}, in Proceedings of the
  nineteenth annual ACM-SIAM symposium on Discrete algorithms, SODA '08,
  Society for Industrial and Applied Mathematics, 2008, pp.~333--342.

\bibitem{C-76}
{\sc Nicos Christofides}, {\em Worst-case analysis of a new heuristic for the
  travelling salesman problem}, tech. report, Carnegie-Mellon Univ. Management
  Sciences Research Group, 1976.

\bibitem{Clarkson99}
{\sc Kenneth~L. Clarkson}, {\em Nearest neighbor queries in metric spaces},
  Discrete Comput. Geom., 22 (1999), pp.~63--93.

\bibitem{CG-06}
{\sc Richard Cole and Lee-Ad Gottlieb}, {\em Searching dynamic point sets in
  spaces with bounded doubling dimension}, in 38th annual ACM symposium on
  Theory of computing, 2006, pp.~574--583.

\bibitem{CL98}
{\sc Artur Czumaj and Andrzej Lingas}, {\em A polynomial time approximation
  scheme for {E}uclidean minimum cost k-connectivity}, in 25th International
  Colloquium on Automata, Languages and Programming, ICALP '98,
  Springer-Verlag, 1998, pp.~682--694.

\bibitem{CLZ02}
{\sc Artur Czumaj, Andrzej Lingas, and Hairong Zhao}, {\em Polynomial-time
  approximation schemes for the {E}uclidean survivable network design problem},
  in 29th International Colloquium on Automata, Languages and Programming,
  ICALP '02, Springer-Verlag, 2002, pp.~973--984.

\bibitem{DFJ-54}
{\sc G.~Dantzig, R.~Fulkerson, and S.~Johnson}, {\em Solution of a large-scale
  traveling-salesman problem}, Operations Research, 2 (1954), pp.~393--410.

\bibitem{FRT-03}
{\sc Jittat Fakcharoenphol, Satish Rao, and Kunal Talwar}, {\em A tight bound
  on approximating arbitrary metrics by tree metrics}, in 35th annual ACM
  symposium on Theory of computing, ACM, 2003, pp.~448--455.

\bibitem{GGN-06}
{\sc Jie Gao, Leonidas~J. Guibas, and An~Nguyen}, {\em Deformable spanners and
  applications}, Comput. Geom. Theory Appl., 35 (2006).

\bibitem{GKK10}
{\sc Lee-Ad Gottlieb, Leonid Kontorovich, and Robert Krauthgamer}, {\em
  Efficient classification for metric data}, in 23rd Conference on Learning
  Theory, Omnipress, 2010, pp.~433--440.

\bibitem{GK11}
{\sc Lee-Ad Gottlieb and Robert Krauthgamer}, {\em A nonlinear approach to
  dimension reduction}, in Proceedings of the Twenty-Second Annual ACM-SIAM
  Symposium on Discrete Algorithms, SODA '11, SIAM, 2011, pp.~888--899.

\bibitem{GR-08}
{\sc Lee-Ad Gottlieb and Liam Roditty}, {\em An optimal dynamic spanner for
  doubling metric spaces}, in Proceedings of the 16th annual European symposium
  on Algorithms, ESA '08, Springer-Verlag, 2008, pp.~478--489.

\bibitem{GKL-03}
{\sc Anupam Gupta, Robert Krauthgamer, and James~R. Lee}, {\em Bounded
  geometries, fractals, and low-distortion embeddings}, in 44th Annual IEEE
  Symposium on Foundations of Computer Science, FOCS '03, IEEE Computer
  Society, 2003, pp.~534--543.

\bibitem{Gutin02}
{\sc Gregory Gutin and Abraham~P. Punnen}, eds., {\em The traveling salesman
  problem and its variations}, vol.~12 of Combinatorial Optimization, Kluwer
  Academic Publishers, Dordrecht, 2002.

\bibitem{K-72}
{\sc Richard Karp}, {\em Reducibility among combinatorial problems}, in
  Complexity of Computer Computations, R.~Miller and J.~Thatcher, eds., Plenum
  Press, 1972.

\bibitem{KR07}
{\sc Stavros~G. Kolliopoulos and Satish Rao}, {\em A nearly linear-time
  approximation scheme for the {E}uclidean $k$-median problem}, SIAM J.
  Comput., 37 (2007), pp.~757--782.

\bibitem{KL-04}
{\sc Robert Krauthgamer and James~R. Lee}, {\em Navigating nets: {S}imple
  algorithms for proximity search}, in 15th Annual ACM-SIAM Symposium on
  Discrete Algorithms, Jan. 2004, pp.~791--801.

\bibitem{Laakso00}
{\sc T.~J. Laakso}, {\em Ahlfors {$Q$}-regular spaces with arbitrary {$Q>1$}
  admitting weak {P}oincar\'e inequality}, Geom. Funct. Anal., 10 (2000),
  pp.~111--123.

\bibitem{Laakso02}
\leavevmode\vrule height 2pt depth -1.6pt width 23pt, {\em Plane with {$A\sb
  \infty$}-weighted metric not bi-{L}ipschitz embeddable to {${\Bbb R}\sp N$}},
  Bull. London Math. Soc., 34 (2002), pp.~667--676.

\bibitem{Lampis12}
{\sc Michael Lampis}, {\em Improved inapproximability for {TSP}}, in
  Approximation, Randomization, and Combinatorial Optimization, vol.~7408 of
  Lecture Notes in Computer Science, Springer, 2012, pp.~243--253.

\bibitem{LP01}
{\sc U.~Lang and C.~Plaut}, {\em Bilipschitz embeddings of metric spaces into
  space forms}, Geom. Dedicata, 87 (2001), pp.~285--307.

\bibitem{LLRS85}
{\sc E.~L. Lawler, J.~K. Lenstra, A.~H. G.~Rinnooy Kan, and D.~B. Shmoys},
  eds., {\em The Traveling Salesman Problem}, Wiley-Interscience series in
  discrete mathematics, 1985.

\bibitem{Lee10blog}
{\sc James~R. Lee}, {\em The {G}\"odel prize, {TSP}, and volume growth (blog
  post)}.
\newblock
  \url{http://tcsmath.wordpress.com/2010/06/24/the-godel-prize-tsp-and-volume-%
growth/ }, June 2010.

\bibitem{M-99}
{\sc Joseph S.~B. Mitchell}, {\em Guillotine subdivisions approximate polygonal
  subdivisions: {A} simple polynomial-time approximation scheme for geometric
  {TSP}, $k$-{MST}, and related problems}, SIAM J. Comput., 28 (1999),
  pp.~1298--1309.

\bibitem{Mitchell07}
{\sc Joseph S.~B. Mitchell}, {\em A {PTAS} for {TSP} with neighborhoods among
  fat regions in the plane}, in 8th Annual ACM-SIAM Symposium on Discrete
  Algorithms, SIAM, 2007, pp.~11--18.

\bibitem{PV-06}
{\sc Christos Papadimitriou and Santosh Vempala}, {\em On the approximability
  of the traveling salesman problem}, Combinatorica, 26 (2006), pp.~101--120.

\bibitem{P-77}
{\sc Christos~H. Papadimitriou}, {\em The {E}uclidean travelling salesman
  problem is {NP}-complete}, Theoretical Computer Science, 4 (1977), pp.~237 --
  244.

\bibitem{PY-93}
{\sc Christos~H. Papadimitriou and Mihalis Yannakakis}, {\em The traveling
  salesman problem with distances one and two}, Math. Oper. Res., 18 (1993),
  pp.~1--11.

\bibitem{RS98}
{\sc Satish~B. Rao and Warren~D. Smith}, {\em Approximating geometrical graphs
  via ``spanners'' and ``banyans''}, in 30th annual ACM symposium on Theory of
  computing, ACM, 1998, pp.~540--550.

\bibitem{Reinelt:TSP}
{\sc G.~Reinelt}, {\em The Traveling Salesman}, vol.~840 of Lecture Notes in
  Computer Science, Springer-Verlag, Berlin, 1994.

\bibitem{S-10}
{\sc Michiel Smid}, {\em On some combinatorial problems in metric spaces of
  bounded doubling dimension}.
\newblock Manuscript, available at
  \url{http://people.scs.carleton.ca/~michiel/research.html}, 2010.

\bibitem{T-04}
{\sc Kunal Talwar}, {\em Bypassing the embedding: algorithms for low
  dimensional metrics}, in 36th annual ACM symposium on Theory of computing,
  ACM, 2004, pp.~281--290.

\bibitem{Trevisan00}
{\sc Luca Trevisan}, {\em When {H}amming meets {E}uclid: {T}he approximability
  of geometric {TSP} and {S}teiner tree}, SIAM Journal on Computing, 30 (2000),
  pp.~475--485.

\end{thebibliography}
}



\end{document}